# Properties of LiMnBO$_3$ glasses and nanostructured glass-ceramics

Przemysław P. Michalski[1,@], Agata Gołębiewska[1], Julien Trébosc[2], Olivier Lafon[2,3],
Tomasz K. Pietrzak[1], Jacek Ryl[4], Jan L. Nowiński[1], Marek Wasiucionek[1], Jerzy E. Garbarczyk[1]

[1] Faculty of Physics, Warsaw University of Technology, Koszykowa 75, 00-662 Warsaw, Poland

[2] Univ. Lille, CNRS, UMR 8181, UCCS-Unité de Catalyse et de Chimie du Solide, F-59000 Lille, France

[3] Institut Universitaire de France (IUF), 1 rue Descartes, 75231 Paris, France

[4] Department of Electrochemistry, Corrosion and Materials Engineering, Faculty of Chemistry, Gdańsk University of Technology, Narutowicza 11/12, 80-233, Gdańsk, Poland

## Abstract

Polycrystalline LiMnBO$_3$ is a promising cathode material for Li-ion batteries. In this work, we investigated the thermal, structural and electrical properties of glassy and nanocrystallized materials having the same chemical composition. The original glass was obtained via a standard melt-quenching method. SEM and $^7$Li solid-state NMR indicate that it contains a mixture of two distinct glassy phases. The results suggest that the electrical conductivity of the glass is dominated by the ionic one. The dc conductivity of initial glass was estimated to be in the order of $10^{-18}$ S.cm$^{-1}$ at room temperature. The thermal nanocrystallization of the glass produces a nanostructured glass-ceramics containing MnBO$_3$ and LiMnBO$_3$ phases. The electric conductivity of this glass-ceramics is increased by 6 orders of magnitude, compared to the starting material at room temperature. Compared to other manganese and borate containing glasses reported in the literature, the conductivity of the nanostructured glass ceramics is higher than that of the previously reported glassy materials. Such improved conductivity stems from the facilitated electronic transport along the grain boundaries.

## 1. Introduction

Polycrystalline LiMnBO$_3$ is considered as a promising cathode material for Li-ion batteries due to several advantages, such as high theoretical gravimetric capacity of up to 222 mAh.g$^{-1}$ as well as abundant and low-cost raw materials [1]. Nevertheless, numerous studies reported poor cycling performances of this material and the capacity rapidly decreases with respect to other materials [1–4]. It has been proposed that the use of nanoparticles can improve the cycling stability and the electric conductivity of cathode material. For example, an average discharge capacity loss of 0.9% per cycle for the first 50 cycles has been reported for nanoparticles of hexagonal LiMnBO$_3$ phase with crystalline size of ca. 15 nm embedded in a matrix of reduced graphite oxide and amorphous carbon [5].

We investigate here the structure, thermal and electrical properties of nanostructured glass-ceramics composite having the same chemical composition as LiMnBO$_3$. Such composite is synthesized by thermal nanocrystallization of glasses. We have recently applied this simple and inexpensive approach to enhance the electrical conductivity of various glass systems e.g. V$_2$O$_5$–P$_2$O$_5$ [6], LiF–V$_2$O$_5$–P$_2$O$_5$ [7], Li$_2$O–FeO–V$_2$O$_5$–P$_2$O$_5$ [8], or boron-containing Li$_2$O–FeO–B$_2$O$_3$ [9].

## 2. Experimental

*Synthesis of glasses*

Amorphous samples of nominal composition 25Li$_2$O·50MnO·25B$_2$O$_3$ (corresponding to LiMnBO$_3$ nominal formula) were prepared from commercial pre-dried chemicals: Li$_2$CO$_3$ (Aldrich, 99.99%), Mn(CH$_3$COO)$_2$·4H$_2$O (Aldrich, >99%) and H$_3$BO$_3$ (POCh – Polish Chemicals, 99.5%), which were ground and mixed in a planetary mill. Then, the powders were melted at 1015 °C for 15 min. To prevent the manganese ions from oxidation, the double crucible method was used [10]. Eventually, the molten mixture was rapidly poured out onto a stainless-steel plate held at the room temperature (RT) and immediately pressed with another plate. Brown-colored, semi-transparent discs with average thickness of ca. 0.7 mm were obtained.

*Characterization of the samples*

Thermal events were observed with a TA Q600 instrument. Differential thermal analysis (DTA) scans were carried out for heating rates 1 °C.min$^{-1}$ and 10 °C.min$^{-1}$ in the temperature range 25–650 °C, in argon flow.

X-ray diffraction (XRD) patterns were collected using a Philips X'Pert Pro apparatus equipped with a Cu anode ($\lambda_{CuK\alpha}$ = 1.542 Å). To study the structural changes with the temperature, Anton-Paar HTK-1200 oven with high-temperature holder was used. The measurements were conducted under nitrogen flow.

The impedance spectroscopy (IS) and conductivity measurements were carried out on a Novocontrol Alpha-A impedance analyzer integrated with an oven and temperature control system. The whole system provided temperature stabilization of 0.1 °C. The frequency range of the ac signal was $10^{-2}$–$10^7$ Hz. Ion-blocking platinum electrodes were sputtered on both sides of the samples. During measurements, the samples were step-wise heated from the RT up to different maximum temperatures in 450–550 °C range and subsequently cooled down to RT. The direct current (dc) resistivity at investigated temperatures was obtained by taking the point, where the phase shift between current and voltage was the smallest. From these values, the dc conductivity was calculated considering geometrical factors.

SEM (scanning electron microscopy)/EDX (energy dispersive X-ray analysis) investigations were performed at the Institute of High Pressure Physics, Polish Academy of Sciences using a Zeiss ULTRA plus microscope with EDS Bruker Quantax 400 system. The images of samples' surface and their elemental compositions were obtained.

Manganese valence states on the surface of analyzed samples were investigated using X-ray photoelectron spectroscopy (XPS), with an Escalab 250Xi spectrometer (Thermofisher Scientific, U.K.), utilizing a monochromatic Al K$\alpha$ source and charge neutralization by means of a flood gun. Spectrometer calibration was conducted on gold crystal prior to the measurements. High-resolution spectra were recorded at a pass energy of 15 eV and an energy step size of 0.1 eV. Avantage analysis software was provided by the manufacturer.

To investigate the local environments of lithium in glasses, Magic Angle Spinning (MAS) solid-state nuclear magnetic resonance measurements were performed at the University of Lille, France, at 1.9 T (i.e. $^1$H Larmor frequency of 100 MHz) on a Bruker Avance NMR spectrometer equipped with 2.5 mm double-resonance HX probe. The rotor was spun at a MAS frequency of 30 kHz. 1D $^7$Li MAS NMR spectra were acquired using rotor-synchronized spin echo sequence ($\pi/2$-$\tau$-$\pi$-$\tau$-acquisition). The $\pi/2$ pulse length was 2 µs. The spectrum results from averaging 16384 transients with a recycle delay of 0.1 s. $^7$Li isotropic chemical shifts were referenced to a 1 M solution of LiCl. The simulation of the $^7$Li spectrum was performed using dmfit software [11].

## 3. Results

*Thermal analysis*

The differential thermal analysis (DTA) scan for as-quenched glass, taken at a heating rate of 10 °C.min$^{-1}$, is presented in Fig. 1. The obtained curve is typical for glassy samples, consisting of a glass transition baseline shift (marked as $T_g$) and crystallization peaks (marked as $T_c$). It is noteworthy that the difference between $T_g$ and $T_{c1}$ temperatures is rather high and equal to 60 °C (compared to 41 °C in LiFeBO$_3$ glass [9]). This indicates a relative resistance of the glass to immediate crystallization once the temperature goes above $T_g$. Also, one can see that the intensity of the glass transition signal is relatively high, as compared to crystallization peaks. Usually, glass transition is not as pronounced as crystallization. However, this may depend on glass system, composition and quenching speed. For example, in case of LiFeBO$_3$ glass [9], the glass transition "step" is roughly half of the height of the crystallization peak. In case of different borate Li$_2$O–Al$_2$O$_3$–MoO$_3$–B$_2$O$_3$ glass system studied by Shaaban *et al.*, the relative height of glass transition "step" varies non-monotonically with MoO$_3$/B$_2$O$_3$ ratio, also being higher than crystallization peak for some compositions [12]. The DTA curve obtained for a slower heating rate of 1 °C.min$^{-1}$ exhibited much less pronounced transitions than in the case of the scan taken at 10 °C.min$^{-1}$: the crystallization peak $T_{c1}$ was not observed and other peaks were not well separated. The corresponding curve is shown in the Supplementary information. The obtained characteristic temperatures for both heating rates are reported in Table 1.

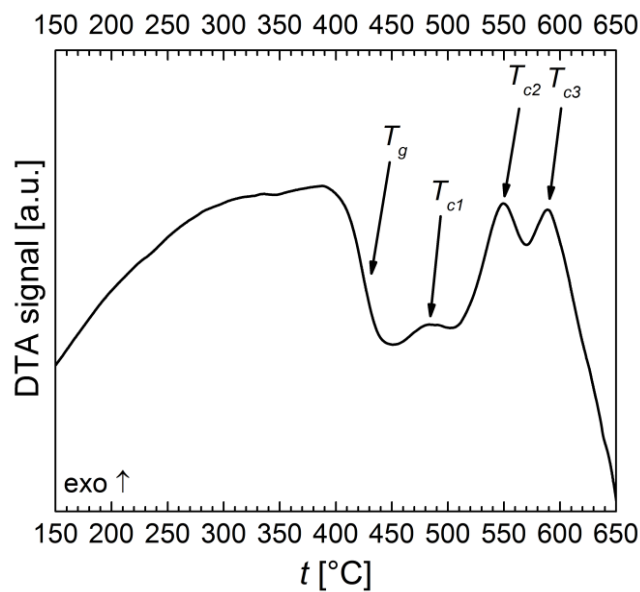

Fig. 1. DTA curve for heating rate 10 °C.min$^{-1}$. Characteristic thermal events were marked with arrows.

Tab. 1. Temperatures of characteristic thermal events for heating rates 10 °C.min$^{-1}$ and 1 °C.min$^{-1}$.

| heating rate / °C.min$^{-1}$ | $T_g$ / °C | $T_{c1}$ / °C | $T_{c2}$ / °C | $T_{c3}$ / °C |
|---|---|---|---|---|
| 10 | 424 | 484 | 550 | 589 |
| 1 | 411 |  | 506 | 568 |

*XRD measurements*

XRD measurements were carried out at temperatures from RT up to 675 °C in order to investigate phases that appeared during annealing. The diffractograms were collected using step-wise heating at ca. 1 h intervals (including heating, temperature stabilization and measurement). During data processing, the background was subtracted from all obtained diffractograms. The results are presented in Fig. 2. The initial sample was fully amorphous. It remains mainly amorphous below 425°C, corresponding to $T_g$. The first diffraction peaks that appeared may be attributed to manganese oxide $Mn_3O_4$ phase (ICDD code: 00-024-0734). At temperatures higher than 500 °C, these phases recrystallized to orthorhombic $MnBO_3$ (o-$MnBO_3$, 475 °C, ICDD code: 00-043-0059), monoclinic $LiMnBO_3$ (m-$LiMnBO_3$, 500 °C, ICDD code: 01-073-4014) and hexagonal $LiMnBO_3$ (h-$LiMnBO_3$, 500 °C, ICDD code: 04-011-8640). These crystallization processes can be associated with $T_{c1}$ temperature. At 575 °C, the peak at 26° matching $LiBO_2$ (ICDD code: 01-076-0578) pattern became visible. This temperature corresponds to $T_{c2}$. After further annealing at higher temperatures (up to 625 °C, exceeding $T_{c3}$) no new peaks may be distinguished, only their intensity increases. After annealing at 625 °C, grain sizes were estimated using Scherrer method. For o-$MnBO_3$, m-$LiMnBO_3$, and h-$LiMnBO_3$ phases, the results were 60–70 nm, 50–55 nm and 45–60 nm, respectively. The cooling down of the sample to the RT only weakly alters the diffractogram. The slight changes in the diffraction angles stems from thermal expansion of the unit cells of the crystal, when lowering the temperature.

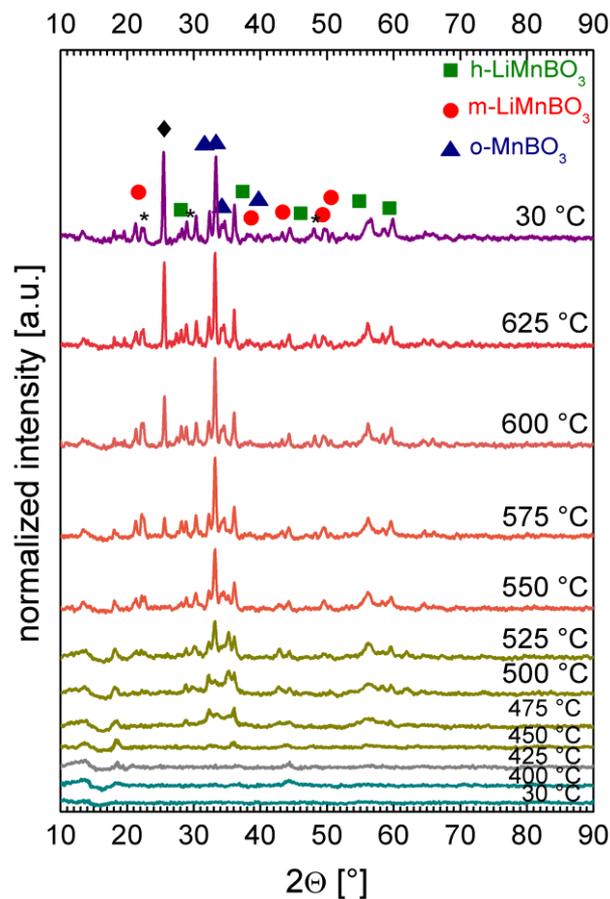

Fig. 2. XRD patterns as function of temperature for step-wise heating. h-$LiMnBO_3$, m-$LiMnBO_3$ and o-$MnBO_3$ patterns are displayed as olive squares, red dots and blue triangles, respectively. $LiBO_2$ peak is marked with diamond and $LiAlSiO_4$ inclusions are marked with asterisks.

We also collected diffractograms after isothermal annealing of 10 and 20 h at 525 °C. The temperature corresponds to that for which the diffraction peaks of $LiMnBO_3$ became clearly visible in Fig. 2. Fig. 3 compares XRD patterns acquired using stepwise and isothermal annealing. The intensity of peaks e.g. at 30°, 34° and 40° ascribed to o-MnBO$_3$, 22°, 33° and 39°, ascribed to m-LiMnBO$_3$ phase, and at 28°, 36° and 55° ascribed to h-LiMnBO$_3$ phase, increases for 10 h annealing time, compared to diffractogram obtained during temperature ramp. The diffractograms obtained after 10 h and 20 h of annealing do not differ from each other, suggesting that 10 h time is enough to perform isothermal nanocrystallization. It can be therefore concluded that the isothermal annealing at temperatures slightly lower than $T_{c2}$ led to further growth of MnBO$_3$ and LiMnBO$_3$ phases. The peaks marked with asterisk may be ascribed to LiAlSiO$_4$ phase which probably crystallized from the impurities introduced by the crucible during the synthesis process (see SEM/EDX section). The peak marked with question mark may not be ascribed to any known crystalline phase. Assuming that the obtained crystalline phases exhibit similar densities and X-ray radiation attenuation factors, one can evaluate the relative weight ratio of those, basing on the peak intensities. The results for the sample annealed for 20 h are presented in Table 2.

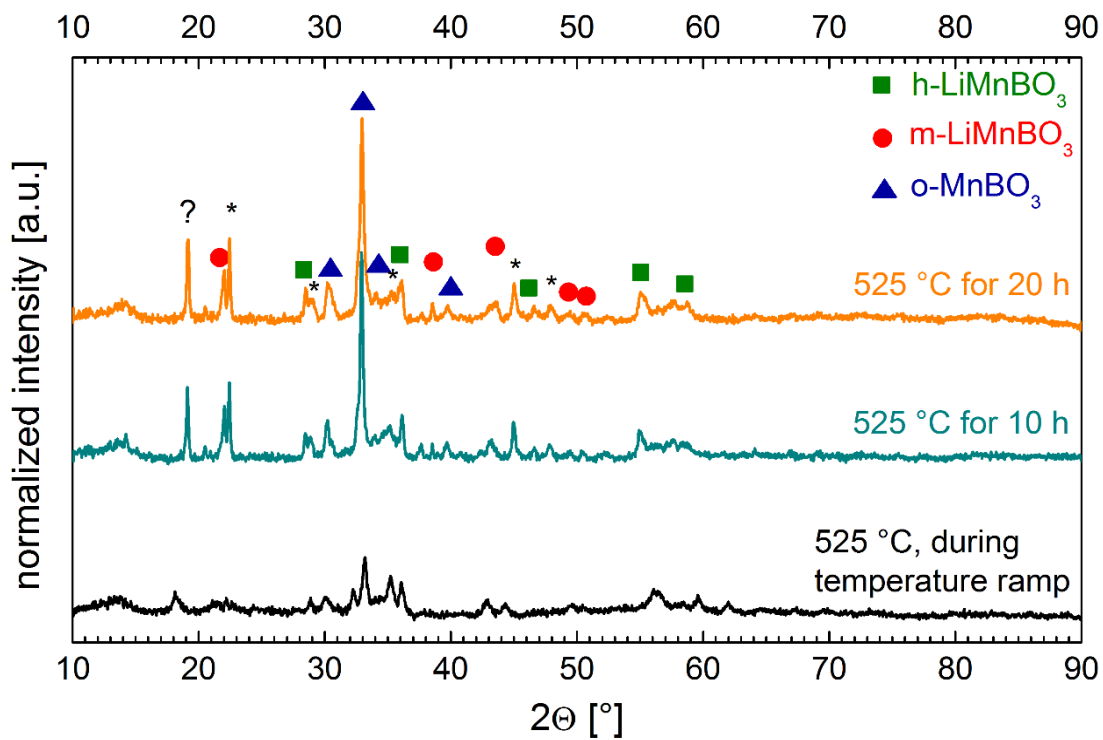

Fig. 3. Comparison of XRD patterns at 525 °C using stepwise heating (bottom) with heating of ca. 1 h at 525 °C and isothermal annealing for 10 h (middle) and 20 h (top). h-LiMnBO$_3$, m-LiMnBO$_3$ and o-MnBO$_3$ patterns are displayed as olive squares, red dots and blue triangles, respectively. LiAlSiO$_4$ inclusions are marked with asterisk, while the peak marked with question mark could not be ascribed to any known phase.

Tab. 2. Relative ratios of different phases for sample isothermally annealed in 525 °C for 20 h.

| [% h-LiMnBO$_3$] | [% m-LiMnBO$_3$] | [% m-MnBO$_3$] | [% MnO$_2$] | [% LiAlSiO$_4$] |
|---|---|---|---|---|
| 15 | 16 | 36 | 13 | 20 |

*Electrical conductivity*

The electrical conductivity σ, was measured, while the glass was heated from RT to 450 °C. Fig. 4 shows that log(σT) of the glass is inversely proportional to the temperature for $T < T_g \approx 410$ °C and hence, can be described by an Arrhenius formula with an activation energy of 1.15 eV. The electrical conductivity of the glass at RT could not be measured directly but was estimated to be equal to 2.6·10$^{-18}$ S.cm$^{-1}$ by extrapolation using Arrhenius formula from the values at high temperatures.

Fig. 4 also displays the data obtained after the annealing of the sample at temperature ranging from 450 to 550 °C, followed by its cooling down to RT. At $T \geq 250$°C, the slope of the curve log(σT) versus $T$ for glass-ceramics is similar to that measured for the glass. Conversely, at $T \leq 175$ °C, log(σT) of the glass-ceramics is still inversely proportional to $T$ but with a lower slope, i.e. a lower activation energy. For instance, after nanocrystallization at 500 °C, the activation energy was lowered more than two times (down to 0.49 eV) with respect to the initial glass and we measured a conductivity for the glass-ceramics at RT of 1×10·10$^{-12}$ S.cm$^{-1}$, a value higher by 6 orders of magnitude than that of the initial glass. The lower activation energy in the glass-ceramics compared to the glass suggests the contribution of electrons to the electrical conduction since the activation energies for ionic conduction are usually higher than that of electronic conduction.

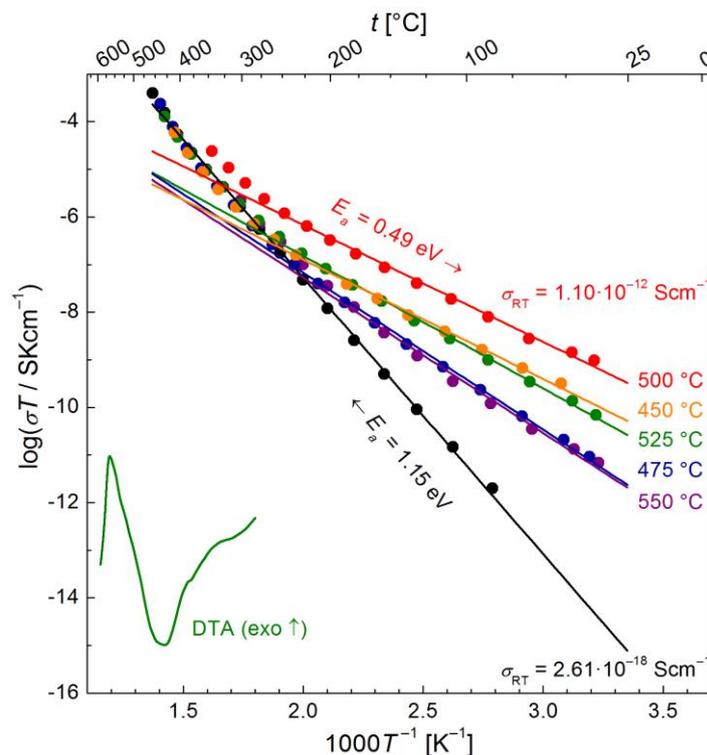

Fig. 4. Arrhenius plot for electrical conductivity. Black dots stand for heating ramp of initial glass, colored dots – cooling ramps after annealing to various nanocrystallization temperatures. DTA curve for 1 °C.min$^{-1}$ is shown for comparison.

*Impedance spectra*

Impedance spectra were acquired to further analyze the charge transport mechanisms. Fig. 5 displays the impedance spectra at 200 °C of the initial glass and glass-ceramics after nanocrystallization at

500 °C. Both impedance spectra contained a low-frequency spur, which is more pronounced in the sample after nanocrystallization. This feature originates from the blocking of lithium ions at the sample-electrode interface. In Fig. 6, the impedance spectrum of the glass-ceramics at 90 °C is presented and exhibits a second semicircle at the low frequencies. Based on the conductivity measurements and impedance spectra, we can postulate that the conductivity in the glass was predominantly ionic. For glass-ceramics, the conductivity was also predominantly ionic in the high temperature range (250–550 °C) but mostly electronic at low temperatures. Further information is provided in Discussion section.

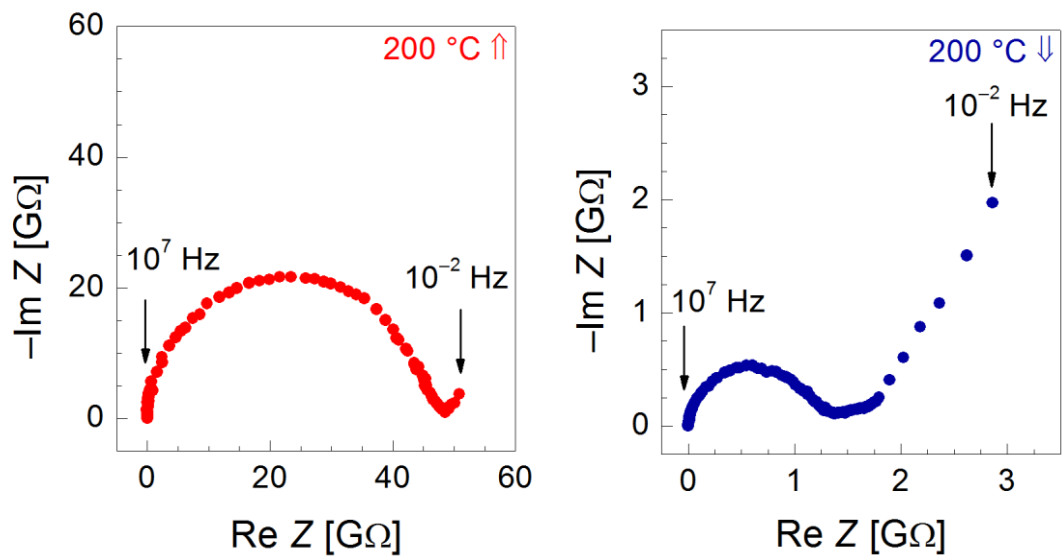

Fig. 5. Impedance spectra at 200 °C of the initial glass (left) during the heating ramp and of a glass-ceramics after nanocrystallization at 500 °C (right), during the cooling ramp.

Fig. 6. Impedance spectrum of sample after nanocrystallization at 500 °C, taken at 90 °C, during cooling ramp.

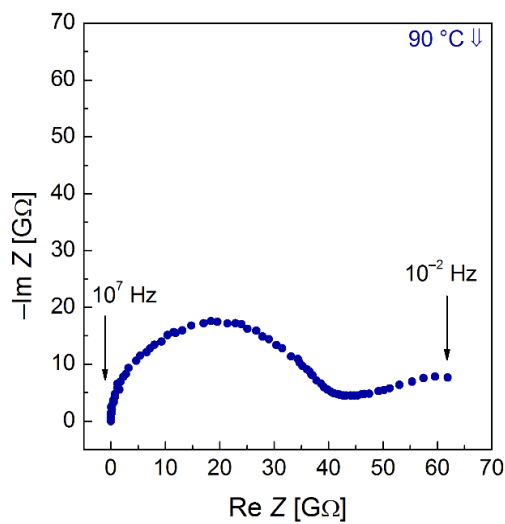

*SEM/EDX investigations*

The morphology of the sample's surface before and after nanocrystallization was observed using scanning electron microscopy. As seen in the Fig. 7a, uniformly distributed, hexagonal shaped bright spots are observed on the darker, smooth surface of the glass. As the diffractograms obtained for the sample (Fig. 2.) did not show any crystalline phases up to 425 °C, those spots must be amorphous or their amount is too low to be detected by XRD. These two domains were separately investigated by EDX and the results for O/Mn composition are presented in Table 3. Light elements, like Li and B cannot be detected using EDX. The bright hexagonal spots are richer in Mn than the nominal composition of the glass, while the darker, smooth surface contains less manganese than the nominal composition. The presence of Al and Si impurities is related to the synthesis process, since the sample melting was carried out in a porcelain crucible, containing Al and Si elements.

Tab. 3. Elemental composition of a glass, as measured by EDX.

| element [atomic %] | hexagons (bright spots) | dark area |
|---|---|---|
| oxygen | 58.5±3.8 | 62.4±5.1 |
| manganese | 30.5±1.7 | 11.4±0.9 |
| impurities (Al+Si) | 7.1±0.2 | 12.0±0.4 |
| O:Mn | 1.9:1 | 5.5:1 |
| O:Mn nominal | 3:1 | |

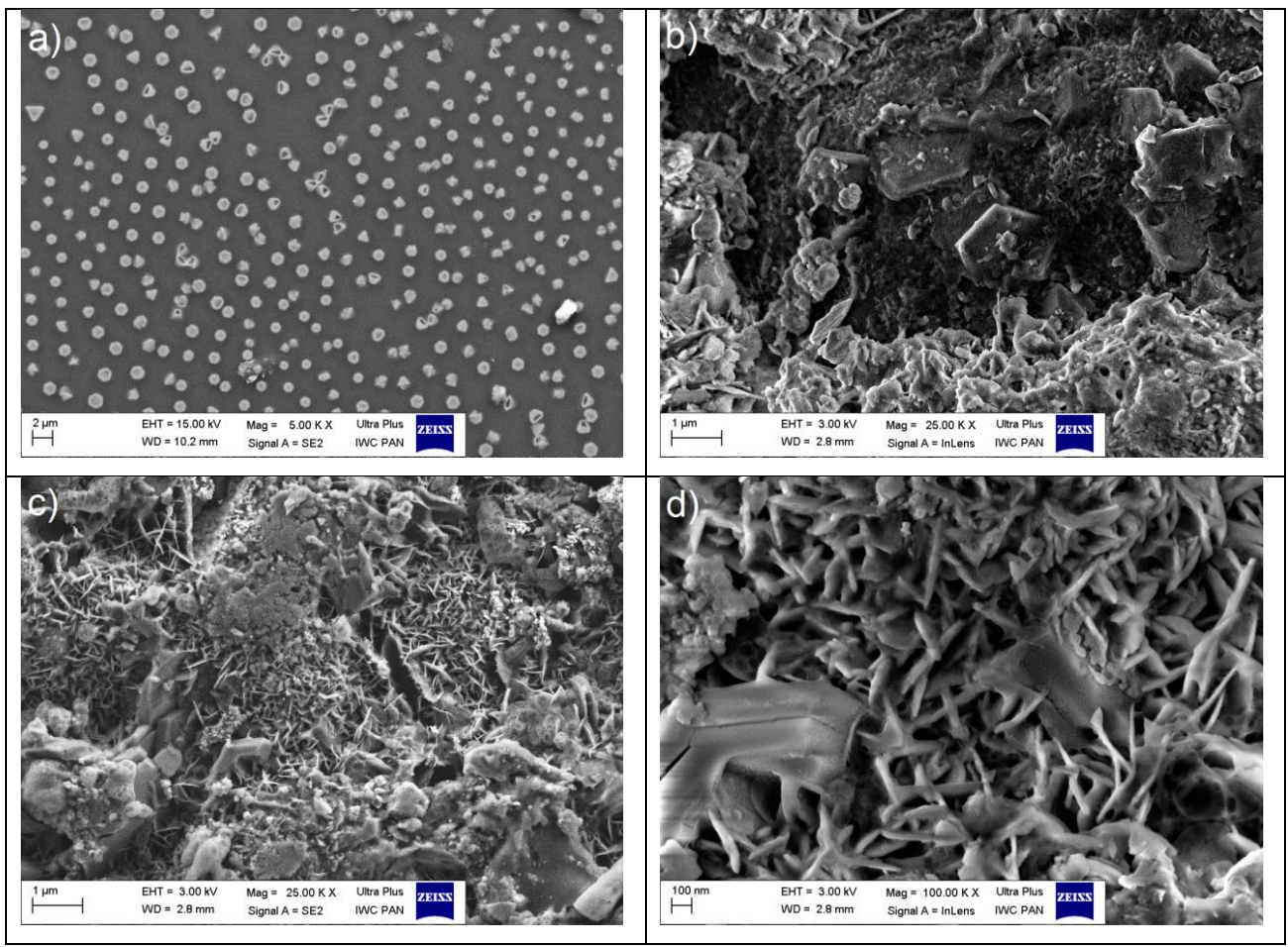

Fig. 7. SEM images of sample surface: a) glass; b–d) glass-ceramics after nanocrystallization at 500 °C.

In the Fig 7b–d, the SEM images for the sample after nanocrystallization at 500 °C are presented.

Generally, one can see the diversity of crystalline structures grown during the nanocrystallization, which are shaped as rods, frozen-like drops, etc. Their main dimension remains lower than 200 nm, which corresponds nicely to values evaluated by XRD. One should consider that by using the Scherrer estimation, only the size of crystalline "core" may be evaluated, while using SEM, the size of a core surrounded by glassy matrix is rather obtained. In the Fig. 7b, the µm-sized hexagonal plates are present, which probably correspond to h-LiMnBO$_3$ phase.

## XPS results

The XPS was used to evaluate the valence states of manganese on the sample's surface. In the Fig. 8, the spectra obtained for glass (a) and nanomaterial (b) with the peaks' deconvolution are shown. The black points depict the experimental data, while the colored and grey lines – the spectra deconvolutions and envelope, respectively.

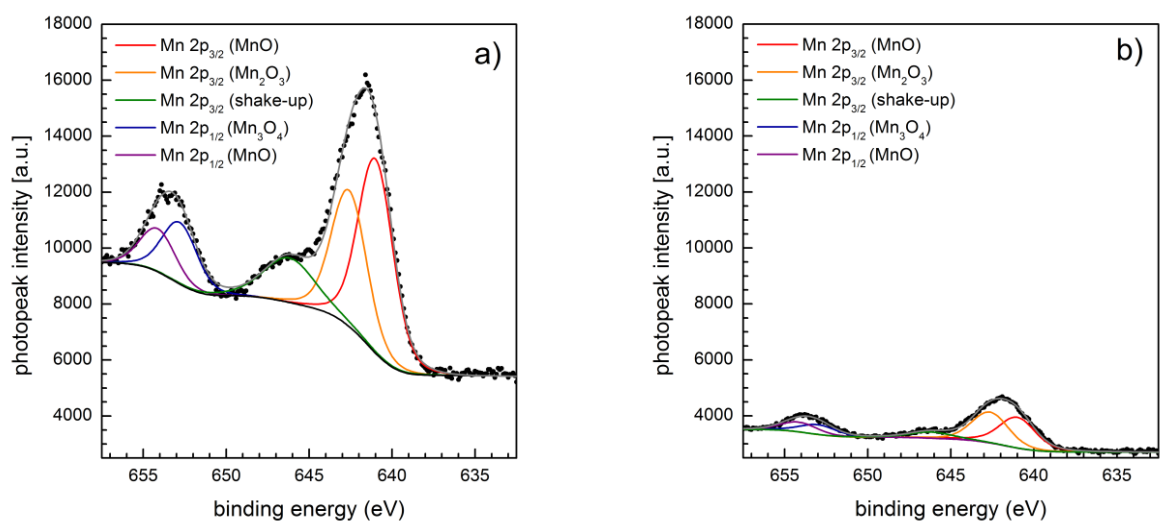

Fig. 8. High-resolution XPS spectra for the glass (left) and for nanomaterial, after annealing at 500 °C (right).

Comparing both spectra, one can see that the manganese signal for nanocrystallized sample is very low. As the XPS is the surface technique, this should be related to manganese diffusion from surface to inner layers, as the glass is annealed. The peaks positions were attributed to proper manganese valence states basing on classical Biesinger [13] and Stranick [14,15] papers and the relative concentration [Mn$^{2+}$]/([Mn$^{2+}$]+[Mn$^{3+}$]) was calculated by considering the peak areas. This ratio equals to 59 and 52% for the glass and nanomaterial, respectively.

## MAS Solid State NMR studies

The experimental 1D $^7$Li MAS NMR spectrum of the glass is shown in Fig. 8. It is dominated by the dipolar coupling between the unpaired electrons and $^7$Li nuclei [16], which can be expressed as a symmetric rank-two tensor and produces spinning sideband patterns analogous to those produced by chemical shift anisotropy. The 1D $^7$Li MAS NMR spectrum was simulated as the sum of spinning sideband manifolds with an anisotropy Δσ and an asymmetry parameter η = 0. The spinning sidebands are represented by Gaussian lineshape of the same full-width at half maximum (FWHM). Two distinct spinning sideband manifolds were necessary and enough to fit the simulated spectrum to the experimental one (see Fig. 9). The best-fit parameters are reported in Table 4. Hence, the lithium ions in glass occupy two distinct environments, which are consistent with the two phases observed by SEM in Fig. 7. Both signals exhibit a significant anisotropy and hence, the $^7$Li nuclei in the two environments are subject to dipolar interaction with the unpaired electrons. The isotropic shifts of the two $^7$Li signals are significantly higher than that (−201 ppm) of monoclinic LiMnBO$_3$ phase [17]. The most deshielded $^7$Li resonance has an isotropic shift close to that (36 ppm) of $^7$Li resonance in orthorhombic LiMnO$_2$ phase and hence, must correspond to similar local environment [18].

Tab. 4. Best-fit parameters for the simulation of 1D $^7$Li MAS NMR spectrum of the initial glass, shown in Fig. 8.

| Fraction / % | δ / ppm | FWHM / ppm | Δσ / kHz |
|---|---|---|---|
| 11.8 | 4.1 | 23.8 | 246 |
| 88.2 | 27.9 | 156.2 | 132 |

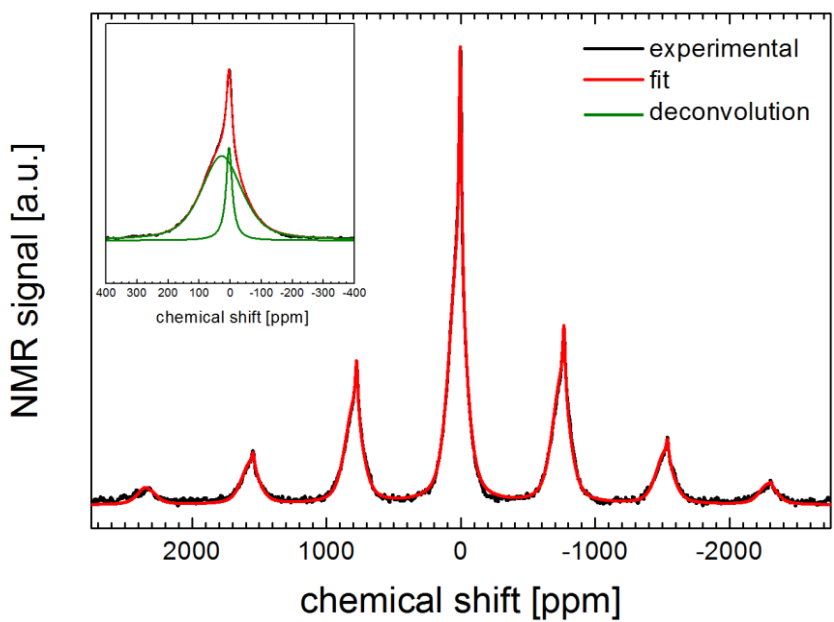

Fig. 9. Experimental (black line) and best-fit simulated (red line) 1D $^7$Li MAS NMR spectrum of the initial glass (black line) at 1.9 T with a MAS frequency of 30 kHz. The inset shows an expansion of the central transition. The simulated $^7$Li spectrum is the sum of two distinct lineshapes (displayed as olive lines in the inset) with parameters given Table 4.

## 4. Discussion

The charge transport by electrons or ions in disordered materials, such as glasses, can be described by Arrhenius formula:

$$\sigma = \frac{\sigma_0}{T} exp\left(-\frac{E_a}{k_B T}\right) \quad (1)$$

Despite the similar formula, the conduction mechanisms are significantly different. Electron transport in glasses containing transition metal ions takes place via polaron hopping between different redox pairs. The expression of the electronic conductivity in disordered materials via such mechanism was given by Mott [19, 20]:

$$\sigma = \nu_{el} c(1-c) \cdot \frac{e^2}{R k_B T} \cdot \exp(-2\alpha R) \exp\left(-\frac{E_a}{k_B T}\right) \quad (2)$$

where $R$ is the average distance between hopping centers, $\nu_{el} \approx \hbar/(mR^2)$ is the optical frequency of polaron hopping with $m$ – effective mass of a polaron, $\alpha$ is the inverse localization length of the electron wave function, $c = [Mn^{2+}]/([Mn^{2+}]+[Mn^{3+}])$ is the fraction of isovalent hopping sites for electrons and $E_a$ is the activation energy of electronic conduction. On the other hand, ionic transport is related to the free volume in the glass and is also enhanced at higher temperatures.

We have postulated that the electrical conductivity of the initial glass is dominated by the ionic component, due to a presence of a "spur" in the impedance spectrum at low frequencies and the high activation energy.

After nanocrystallization, the ionic conductivity was still predominant in the high temperature region (550–250 °C). However, we measured a gradual reduction in the activation energy in the intermediate temperatures (250–175 °C). Such reduction suggests the larger contribution of electrons to the electrical conductivity of glass-ceramics for decreasing temperature, since the activation energy of electronic conductivity is lower than that of ionic one. For $T \leq 175$°C, the electronic conductivity prevails and the activation energy is reduced to 0.49 eV, instead of 1.15 eV for ionic conductivity. Similar values were obtained e.g. for purely electronic conducting $FeO$–$P_2O_5$ glassy system [21]. Furthermore, at low temperature, the impedance spectra of the glass-ceramics consist of two semicircles – which may be interpreted as an appearance of the electronic conduction path "shunting" the ionically conducting system. Conduction via electrons in the glass-ceramics is possible owing to significant change in microstructure with respect to the initial glass. The glass-ceramics contains highly defected boundaries between the grain and the glassy phases. These boundaries, which contains a larger number of $Mn^{2+}/Mn^{3+}$ redox pairs, serve as paths of facilitated electronic transport. The XPS measurements revealed that, after nanocrystallization, the relative concentration of $Mn^{2+}$ ions is approximately 50%, meaning that the $c$ parameter is close to optimum.

The low conductivity of manganese-containing glasses is a common problem. The conductivity of glass-ceramics obtained in this work is one of the highest reported in literature, when compared to other glassy materials. Veeranna Gowda *et al.* measured dc conductivities of lithium-manganese-borate glasses with various $Li_2O$:$MnO_2$:$B_2O_3$ ratios ranging from $2\times10^{-16}$ to $9\times10^{-11}$ S.cm$^{-1}$ at RT.

The measured activation energies ranged from 0.77 eV to 1.12 eV and were comparable to those reported here. Nevertheless, the impedance spectra consisted of one or two semicircles, without the presence of ionic spurs. Despite that, authors concluded that the conductivity is dominated by $Li^+$ ions, as variation of $MnO_2$ content resulted in only minor changes of the values [22]. Min *et al.* studied glasses of composition $xLiMn_2O_4 \cdot (100–x)B_2O_3$ ($x$ = 30, 40, 50) and obtained dc conductivity values $6.2 \times 10^{-19}$ to $6.1 \times 10^{-15}$ S.cm$^{-1}$ at RT [23]. Kupracz *et al.* studied the properties of $MnO–SiO_2–B_2O_3$ glasses with pure electronic conduction [24, 25]. The dc conductivity of the glasses at RT was in the range $10^{-19}$–$10^{-14}$ S.cm$^{-1}$ and the activation energy values ranged from 0.86 to 1.25 eV, depending on composition. The authors concluded that the electronic transport in manganese glasses may be explained by the means of large polaron hopping. The polaron radius was, however, always lower than the mean Mn–Mn distance (even for 0.6 mole Mn) which may partly explain the low conductivity of manganese glasses. On the other hand, Ragupathi *et al.* studied the properties of crystalline $LiMnBO_3$ with additional glassy phase of $LiBO_2$ and obtained the bulk conductivity value equal to $3.6 \times 10^{-7}$ S.cm$^{-1}$ [26]. It should be, however, pointed that the material was prepared using sol-gel method which require the use of chelating medium (in this case: propionic acid, $C_3H_6O_2$), that also act as a carbon source. This additional carbon may improve the conductivity. In our work (and in other works cited above), no possible carbon source was used and the conductivity enhancement is related to thermal nanocrystallization.

SEM images revealed the presence of two regions on the surface of the glass: manganese-rich (bright) particles and manganese-poor background (dark). Such phases separation is well recognized in the literature for the glasses containing a high amount of manganese (e.g. [20, 24, 25]). Typically, this separation took place only when no alkali metal was added to the glass. In our case, the situation is different and the possible explanation may be related to the presence of Al and Si impurities, introduced during melting process. NMR measurements showed also the presence of two lithium local environments, which may correspond to Li nuclei in the two phases.

## 5. Conclusions

In this research, a glass with nominal composition $25Li_2O \cdot 50MnO \cdot 25B_2O_3$ identical to that of $LiMnBO_3$ crystal has been prepared using conventional melt-quenching method. Its amorphousness was confirmed by DTA and XRD measurements. The dc conductivity of initial glass was estimated to be in the order of $10^{-18}$ S.cm$^{-1}$ at room temperature. The SEM images showed the presence of two phases, a manganese-rich one and a manganese-poor one. After nanocrystallization, the desired $MnBO_3$ and $LiMnBO_3$ phases were obtained. Nanocrystallization led to irreversible conductivity increase by 4–6 orders of magnitude. The collected data suggest that the conductivity in glass is dominated by the ionic component in the whole temperature range, whereas, after thermal nanocrystallization, the ionic conductivity only prevails at high temperatures but at low temperature, the conduction is ensured by electrons.

## Acknowledgments

This project has received funding from the European Union's Horizon 2020 research and innovation program under grant agreement No 731019 (EUSMI).

## Other information

Declarations of interest: none.

## References

[1] S. Afyon, D. Kundu, F. Krumeich, R. Nesper, *Nano LiMnBO$_3$, a high-capacity material for Li-ion batteries*. Journal of Power Sources **224** (2013), 145–151.

[2] J.C. Kim, C.J. Moore, B. Kang, G. Hautier, A. Jain, G. Ceder, *Synthesis and Electrochemical Properties of Monoclinic LiMnBO$_3$ as a Li Intercalation Material*. Journal of the Electrochemical Society **158** (2011), A309–A315. DOI: https://doi.org/10.1016/j.jpowsour.2012.09.099

[3] K.-J. Lee, L.-S. Kang, S. Uhm, J.S. Yoon, D.-W. Kim, H.S. Hong, *Synthesis and characterization of LiMnBO$_3$ cathode material for lithium ion batteries*. Current Applied Physics **13** (2013), 1440–1443. DOI: https://doi.org/10.1016/j.cap.2013.04.027

[4] Y.-S. Lee, H. Lee, *Electrochemical properties of LiMnBO$_3$ as a potential cathode material for lithium batteries*. Journal of Ceramic Processing Research **13-S2** (2012), 237–240.

[5] S. Afyon, D. Kundu, A. Darbandi, H. Hahn, F. Krumeich, R. Nesper, *A low dimensional composite of hexagonal lithium manganese borate (LiMnBO$_3$), a cathode material for Li-ion batteries*. Journal of Materials Chemistry A **2** (2014), 18946–18951. DOI: https://doi.org/ 10.1039/C4TA04209C

[6] T.K. Pietrzak, J.E. Garbarczyk, M. Wasiucionek, J.L. Nowiński, *Nanocrystallization in vanadate-phosphate and lithium-iron-vanadate-phosphate glasses*. Physics and Chemistry of Glasses: European Journal of Glass Science **57** (2016), 113–124. DOI: https://doi.org/10.13036/17533562.57.3.038

[7] T.K. Pietrzak, P.P. Michalski, M. Wasiucionek, J.E. Garbarczyk, *Synthesis of nanostructured Li$_3$Me$_2$(PO$_4$)$_2$F$_3$ glass-ceramics (Me = V, Fe, Ti)*. Solid State Ionics **288** (2016), 193–198. DOI: https://doi.org/10.1016/j.ssi.2015.11.021

[8] T.K. Pietrzak, M. Wasiucionek, I. Gorzkowska, J.L. Nowiński, J.E. Garbarczyk, *Novel vanadium-doped olivine-like nanomaterials with high electronic conductivity*. Solid State Ionics **251** (2013) 40–46. DOI: https://doi.org/10.1016/j.ssi.2013.02.012

[9] P.P. Michalski, T.K. Pietrzak, J.L. Nowiński, M. Wasiucionek, J.E. Garbarczyk, *Novel*

nanocrystalline mixed conductors based on LiFeBO$_3$ glass. Solid State Ionics **302** (2017), 40–44. DOI: https://doi.org/10.1016/j.ssi.2016.12.002

[10] K. Hirose, T. Honma, Y. Benino, T. Komatsu, *Glass–ceramics with LiFePO$_4$ crystals and crystal line patterning in glass by YAG laser irradiation*. Solid State Ionics **178** (2007), 801–807. DOI: https://doi.org/10.1016/j.ssi.2007.03.003

[11] D. Massiot, F. Fayon, M. Capron, I. King, S. Le Calvé, B. Alonso, J.-O. Durand, B. Bujoli, Z Gan, G. Hoatson, *Modelling one- and two-dimensional solid-state NMR spectra*. Magnetic Resonance in Chemistry **40** (2002), 70–76. DOI: https://doi.org/10.1002/mrc.984

[12] KH. S. Shaaban, S.M. Abo-naf, A.M. Abd Elnaeim, M.E.M. Hassouna, *Studying effect of MoO3 on elastic and crystallization behavior of lithium diborate glasses*. Applied Physics A **123** (2017), 459. DOI: https://doi.org/10.1007/s00339-017-1052-9

[13] M.C. Biesinger, B.P. Payne, A.P. Grosvenor, L.W.M. Lau, A.R. Gerson, R.St.C. Smart, *Resolving surface chemical states in XPS analysis of first row transition metals, oxides and hydroxides: Cr, Mn, Fe, Co and Ni*. Applied Surface Science **257** (2011), 2717–2730. DOI: https://doi.org/10.1016/j.apsusc.2010.10.051

[14] M.A. Stranick, *MnO$_2$ by XPS*. Surface Science Spectra **6** (1999), 31–48. DOI: https://doi.org/10.1116/1.1247888

[15] M.A. Stranick, *Mn$_2$O$_3$ by XPS*. Surface Science Spectra **6** (1999), 39–46. DOI: https://doi.org/10.1116/1.1247889

[16] C.P. Grey, N. Dupré, *NMR Studies of Cathode Materials for Lithium-Ion Rechargeable Batteries*. Chemical Reviews **104** (2004), 4493–4512. DOI: https://doi.org/ 10.1021/cr020734p

[17] J. C. Kim, X. Li, C.J. Moore, S.-H. Bo, P.G. Khalifah, C.P. Grey, G. Ceder, *Analysis of Charged State Stability for Monoclinic LiMnBO$_3$ Cathode*. Chemistry of Materials **26** (2014), 4200–4206. DOI: http://doi.org/ 10.1021/cm5014174

[18] C.P. Grey, Y. J. Lee, *Lithium MAS NMR studies of cathode materials for lithium-ion batteries*. Solid State Sciences **5** (2003), 883–894. DOI: https://doi.org/10.1016/S1293-2558(03)00113-4

[19] I. G. Austin, N. F. Mott, *Polarons in crystalline and non-crystalline materials*. Advances in Physics **18** (1969) 41–102. DOI: https://doi.org/10.1080/00018736900101267

[20] N.F. Mott, *Electrons in disordered structures*. Advances in Physics **16** (1967), 49–144. DOI: https://doi.org/10.1080/00018736700101265

[21] T.K. Pietrzak, Ł. Wewiór, J.E. Garbarczyk, M. Wasiucionek, I. Gorzkowska, J.L. Nowiński, S. Gierlotka, *Electrical properties and thermal stability of FePO$_4$ glasses and nanomaterials*. Solid State Ionics **188** (2011), 99–103. DOI: https://doi.org/10.1016/j.ssi.2010.11.006

[22] V.C. Veeranna Gowda, R.V. Anavekar, *Transport properties of Li$_2$O–MnO$_2$–B$_2$O$_3$ glasses*. Solid State Ionics **176** (2005), 1393–1401. DOI: https://doi.org/10.1016/j.ssi.2005.04.002

[23] J. Min, L. Chen, J. Wang, R. Xue, W. Cui, *Electronic conductivity of LiMn$_2$O$_4$–B$_2$O$_3$ and LiMn$_2$O$_4$–B$_2$O$_3$–P$_2$O$_5$ glasses*. physica status solidi (a) **146** (1994), 771–776. DOI: https://doi.org/10.1002/pssa.2211460222

[24] P. Kupracz, J. Karczewski, M. Prześniak-Welenc, N.A. Szreder, M.J. Winiarski, T. Klimczuk, R.J. Barczyński, *Microstructure and electrical properties of manganese borosilicate glasses*. Journal of Non-Crystalline Solids **423–424** (2015), 68–75. DOI: https://doi.org/10.1016/j.jnoncrysol.2015.05.014

[25] P. Kupracz, A. Lenarciak, M. Łapiński, M. Prześniak-Welenc, N.A. Wójcik, R.J. Barczyński, *Polaron hopping conduction in manganese borosilicate glass*. Journal of Non-Crystalline Solids **458** (2017), 15–21. DOI: https://doi.org/10.1016/j.jnoncrysol.2016.12.008

[26] V. Ragupathi, M. Safiq, P. Panigrahi, T. Hussain, S. Raman, R. Ahuja, G.S. Nagarajan, *Enhanced electrochemical performance of LiMnBO$_3$ with conductive glassy phase: a prospective cathode material for lithium-ion battery*. Ionics 23 (2017), 1645–1653. DOI: https://doi.org/10.1007/s11581-017-2019-8

**Table captions**

Tab. 1. Temperatures of characteristic thermal events for heating rates 10 °C.min$^{-1}$ and 1 °C.min$^{-1}$.

Tab. 2. Relative ratios of different phases for sample isothermally annealed in 525 °C for 20 h.

Tab. 3. Elemental composition of a glass, as measured by EDX.

Tab. 4. Best-fit parameters for the simulation of 1D $^7$Li MAS NMR spectrum of the initial glass, shown in Fig. 8.

**Figure captions**

Fig. 1. DTA curve for heating rate 10 °C.min$^{-1}$. Characteristic thermal events were marked with arrows.

Fig. 2. XRD patterns as function of temperature for step-wise heating. h-LiMnBO$_3$, m-LiMnBO$_3$ and o-MnBO$_3$ patterns are displayed as olive squares, red dots and blue triangles, respectively. LiBO$_2$ peak is marked with diamond and LiAlSiO$_4$ inclusions are marked with asterisks.

Fig. 3. Comparison of XRD patterns at 525 °C using stepwise heating (bottom) with heating of ca. 1 h at 525 °C and isothermal annealing for 10 h (middle) and 20 h (top). h-LiMnBO$_3$, m-LiMnBO$_3$ and o-MnBO$_3$ patterns are displayed as olive squares, red dots and blue triangles, respectively. LiAlSiO$_4$ inclusions are marked with asterisk, while the peak marked with question mark could not be ascribed to any known phase.

Fig. 4. Arrhenius plot for electrical conductivity. Black dots stand for heating ramp of initial glass, colored dots – cooling ramps after annealing to various nanocrystallization temperatures. DTA curve

for 1 °C.min$^{-1}$ is shown for comparison.

Fig. 5. Impedance spectra at 200 °C of the initial glass (left) during the heating ramp and of a glass-ceramics after nanocrystallization at 500 °C (right), during the cooling ramp.

Fig. 6. Impedance spectrum of sample after nanocrystallization at 500 °C, taken at 90 °C, during cooling ramp.

Fig. 7. SEM images of sample surface: a) glass; b–d) glass-ceramics after nanocrystallization at 500 °C.

Fig. 8. High-resolution XPS spectra for the glass (left) and for nanomaterial, after annealing at 500 °C (right).

Fig. 9. Experimental (black line) and best-fit simulated (red line) 1D $^7$Li MAS NMR spectrum of the initial glass (black line) at 1.9 T with a MAS frequency of 30 kHz. The inset shows an expansion of the central transition. The simulated $^7$Li spectrum is the sum of two distinct lineshapes (displayed as olive lines in the inset) with parameters given Table 4.

# Supplementary information

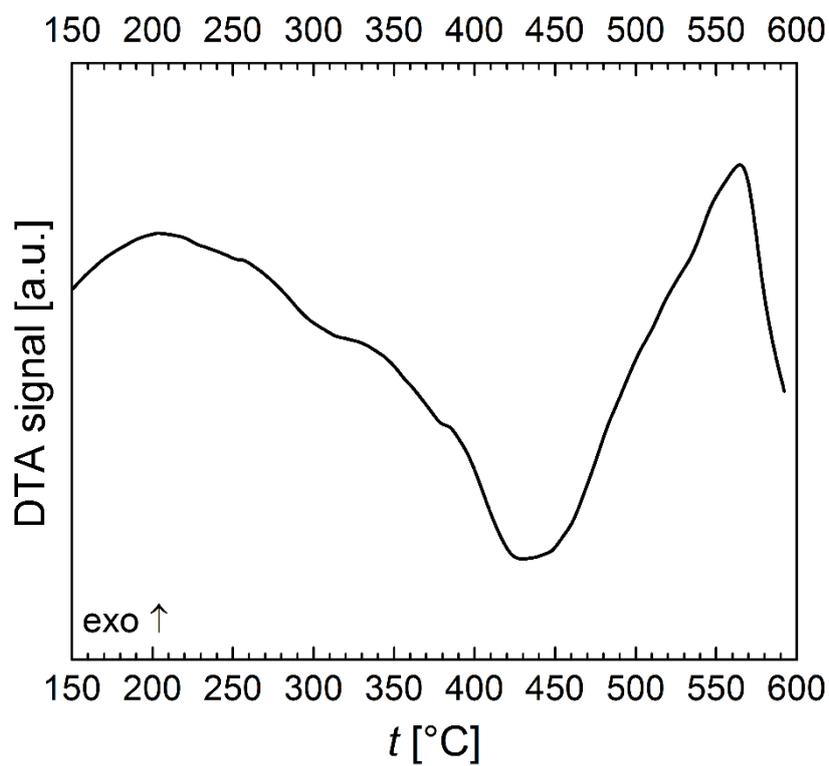

Fig. S1: DTA curve for heating rate 1 °C.min$^{-1}$.

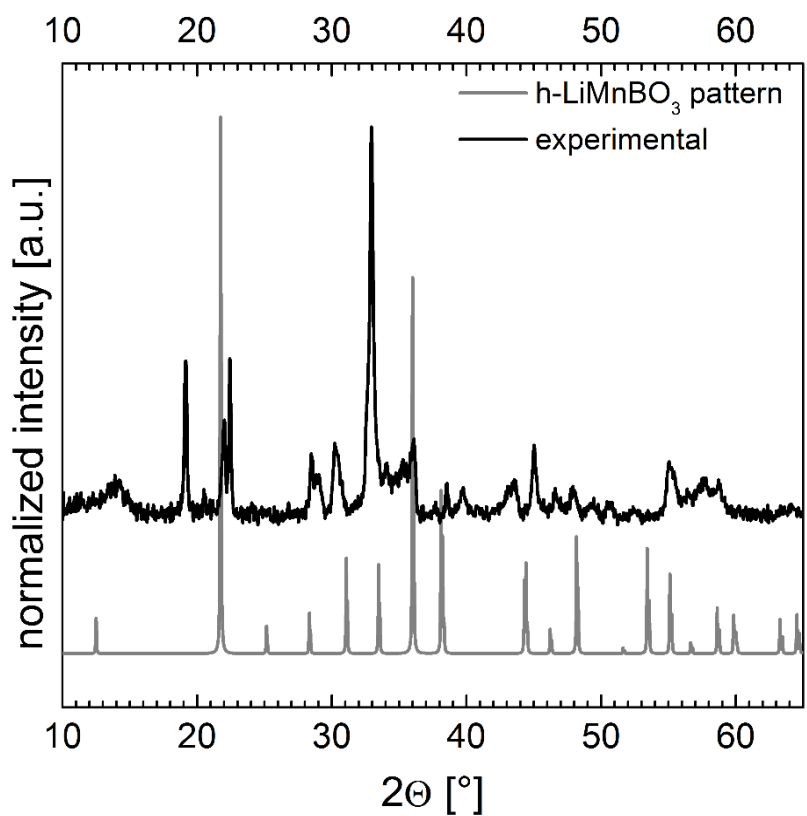

Fig. S2: Comparison of XRD pattern taken after 20 h of annealing at 525 °C (black) with h-LiMnBO$_3$ simulated pattern (grey).

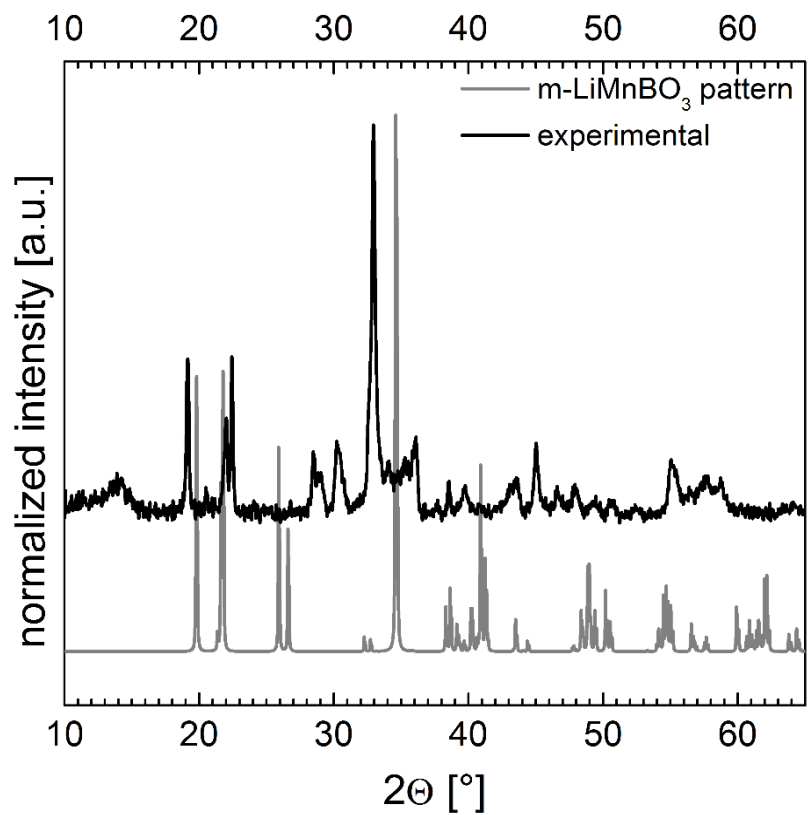

Fig. S3: Comparison of XRD pattern taken after 20 h of annealing at 525 °C (black) with m-LiMnBO$_3$ simulated pattern (grey).

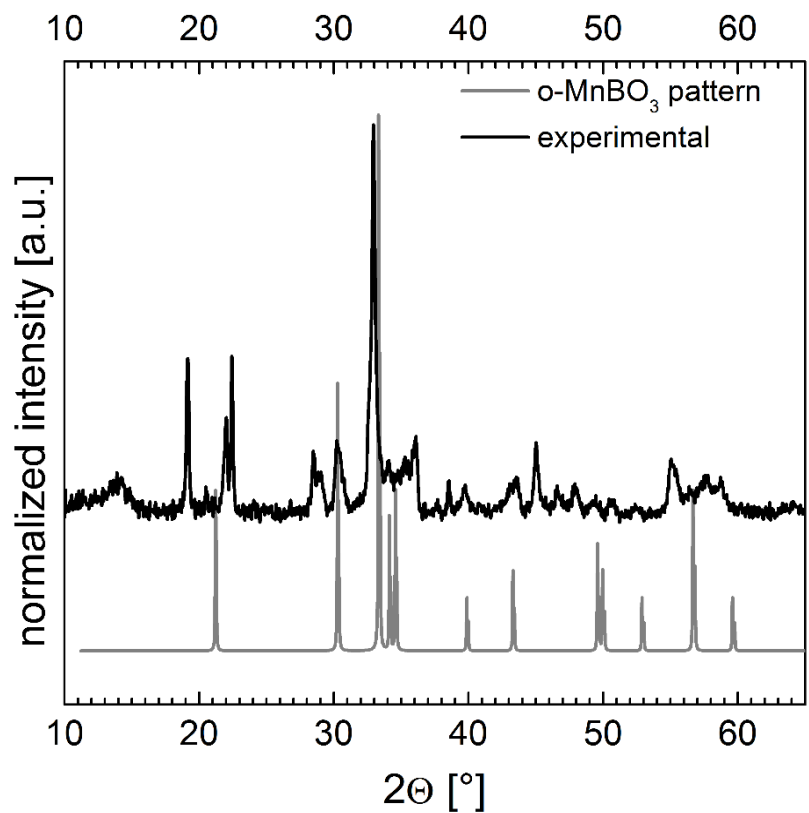

Fig. S4: Comparison of XRD pattern taken after 20 h of annealing at 525 °C (black) with o-MnBO$_3$ simulated pattern (grey).

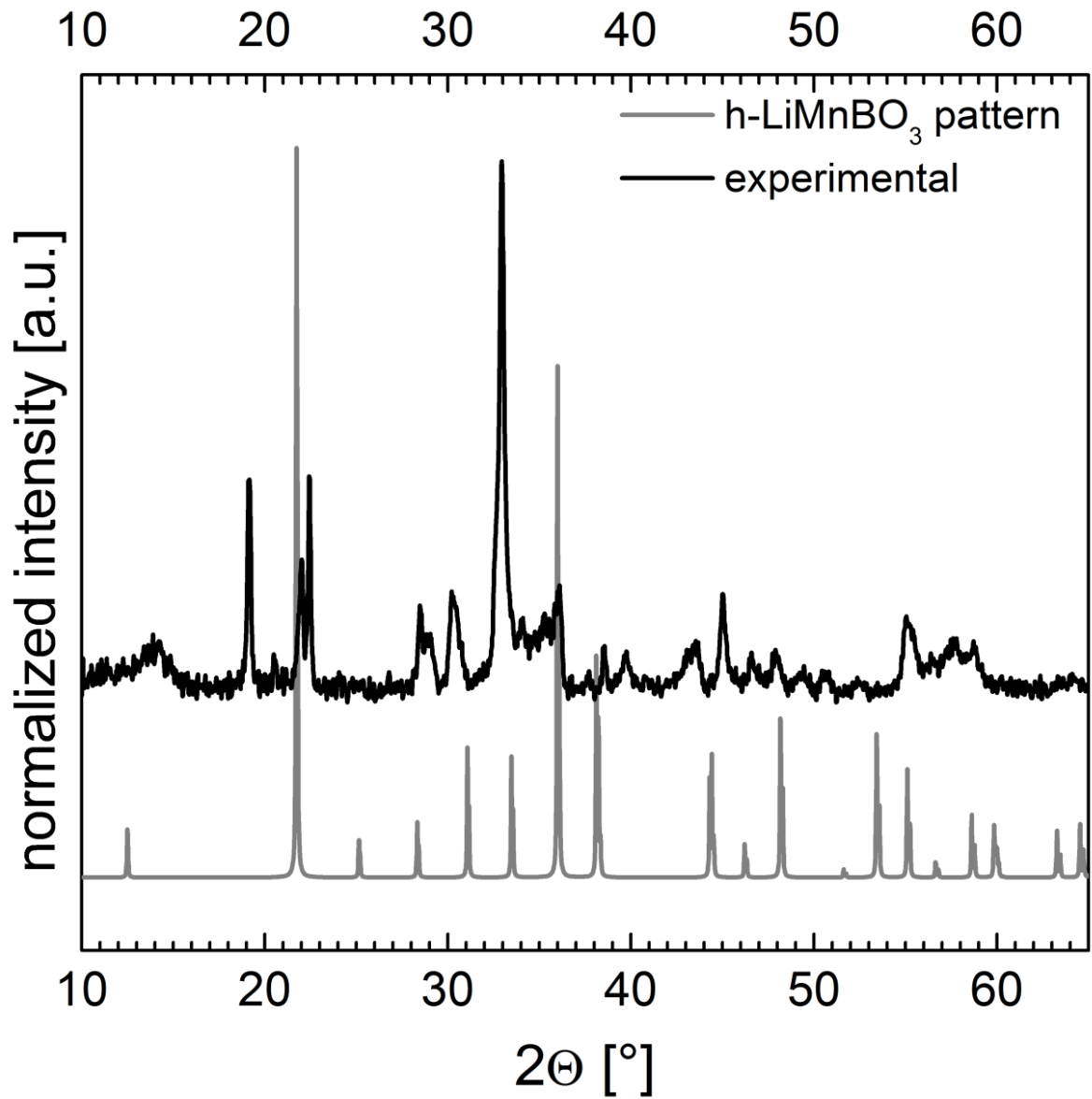

Fig. S1: Comparison of XRD pattern taken after 20 h of annealing at 525 °C (black) with h-LiMnBO$_3$ simulated pattern (grey).

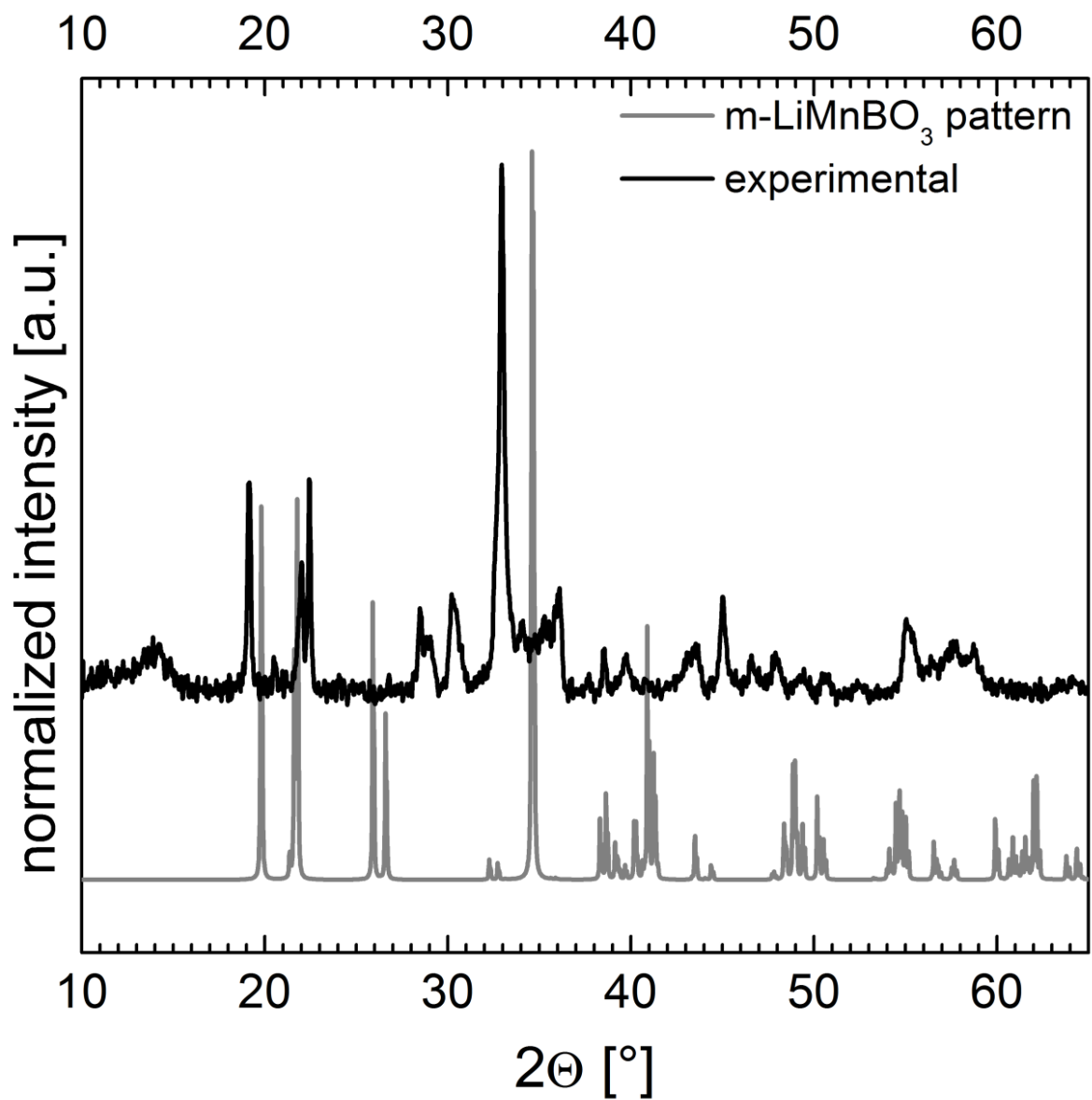

Fig. S2: Comparison of XRD pattern taken after 20 h of annealing at 525 °C (black) with m-LiMnBO$_3$ simulated pattern (grey).

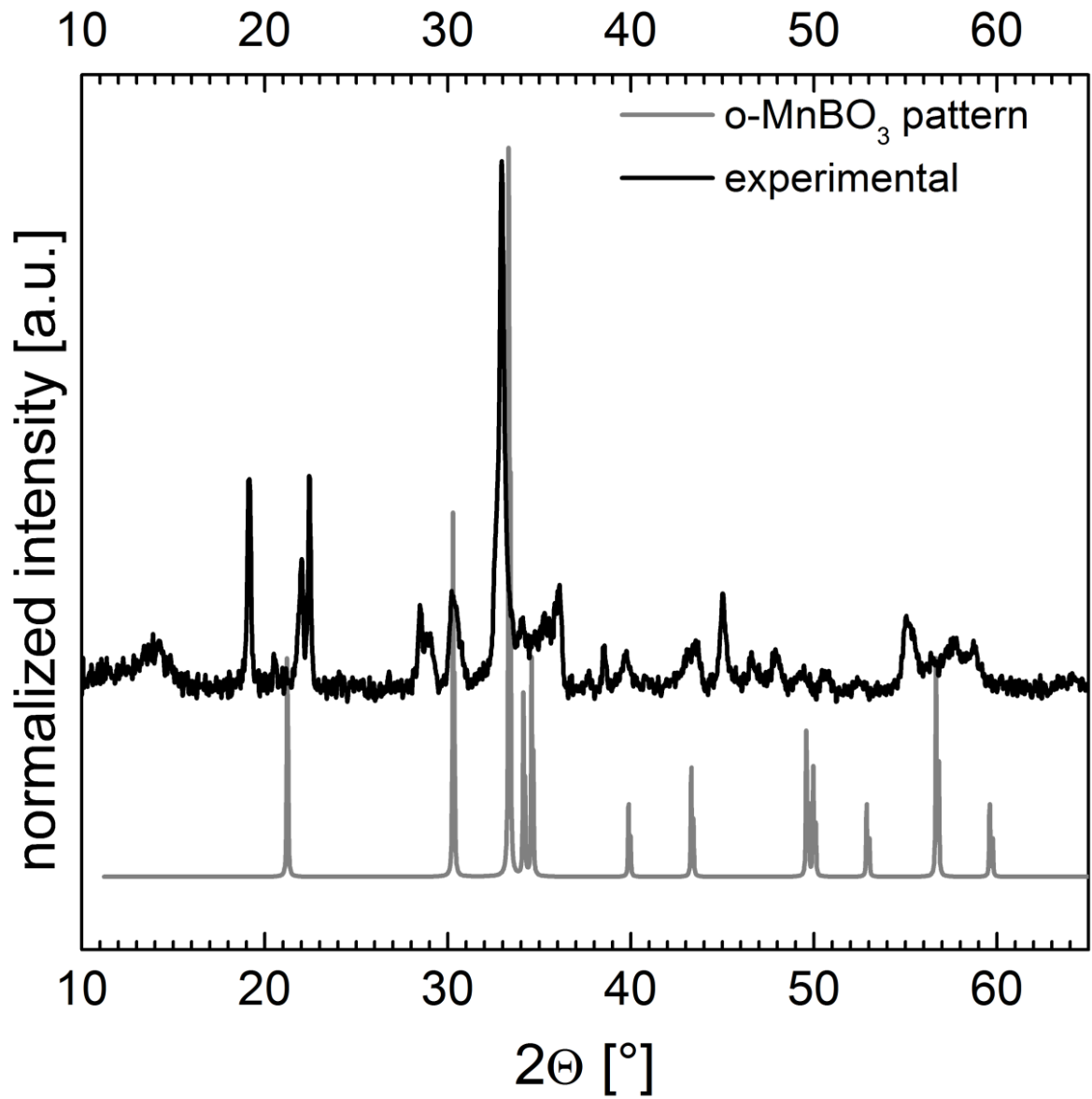

Fig. S3: Comparison of XRD pattern taken after 20 h of annealing at 525 °C (black) with o-MnBO$_3$ simulated pattern (grey).